\documentstyle[twocolumn,prl,aps]{revtex}
\input psfig
\begin{document}
\twocolumn[\hsize\textwidth\columnwidth\hsize\csname @twocolumnfalse\endcsname
\draft
\title{Charge occupancy of two 
interacting electrons on artificial molecules -- exact
results}
\author{A. Aharony}
\address{\it School of Physics and Astronomy, Raymond and
Beverly Sackler
Faculty of Exact Sciences, \\
Tel Aviv University, Tel Aviv 69978, Israel \\ }
\author{O. Entin-Wohlman\cite{OEW}}
\address{Centre for Advanced Studies, The Norwegian Academy, Oslo
0271, Norway}
\author
{Y. Imry and Y. Levinson}
\address{\it Condensed Matter Physics, The Weizmann Institute of
Science, Rehovot 76100, Israel}
\date{\today}
\maketitle

\begin{abstract}
We present exact solutions for two interacting electrons
 on an artificial atom
and on an artificial molecule
made by one and two (single level) quantum dots connected by ideal leads.
Specifically, we calculate the accumulated charge on the dots
as function of the gate voltage $\epsilon_0$, for various strengths of
the
electron-electron interaction $U$ and of the hybridization between
the dots and the (one-dimensional) leads $\gamma$.
For $\gamma<1$ and $2(\gamma-1) \equiv \epsilon_{00}<\epsilon_0<0$, 
there are no bound states.
As $\epsilon_0$ decreases beyond $\epsilon_{00}$, the
accumulated charge $P$ in the two-electron ground state
increases in gradual steps 
from 0 to 1 and then to 2. The values $P \sim 0$ represent
an ``insulating" state, where both electrons are bound to shallow states on
the impurities. The value of $P \approx 1$
corresponds to a ``metal", with one electron localized on the
dots
and the other extended on the leads. The value of 2 corresponds
to another  ``insulator", with both
electrons strongly localized. 
The width of the ``metallic" regime diverges with
$U$
for the single dot, but remains very narrow for the double dot. These
results are contrasted with the simple Coulomb blockade picture.
\end{abstract}

]


\section{Introduction}

Much of the early work on quantum dots concentrated on
large dots, which may contain many electrons,
in the limit of very weak coupling
between the dots and the leads, where one can employ
the simple Coulomb blockade picture.
\cite{CB} In that limit, the energy cost
of adding a single electron to the dot is of order of the
excess energy for the added charge. For a total of two electrons, this is 
of the order of the
Coulomb repulsion between the two electrons.
One expects that for a good contact between the dots and the leads this
energy cost will be reduced. However, the
study of a general coupling strength presents a great challenge.
The case of an `open' dot, connected by quantum point
contacts to a two-dimensional electron gas, \cite{K} has been analyzed
using bozonization techniques and mapping of the Hamiltonian onto the
two- and
four-channel Kondo problem. \cite{M}

It has only recently become possible to also study small quantum dots,
which have a small number of states and contain a small number of
electrons. \cite{zhitenev,tarucha}
Such a small dot, connected to external leads, is similar to a donor
in a doped semiconductor: both may be modeled as an
`impurity' connected to external leads. \cite{ng}
A set of such quantum dots, or an artificial molecule, can then be
modeled by a set of such `impurities'.
In what follows, we sometimes interchange
the terms `dot' and `impurity'.
It is usually assumed that the electrons
interact with each other only when they are on the same quantum dot,
and behave as free electrons when they are on the leads. In what follows
we therefore assume a contact interaction, which exists only on the
dots, and present exact results for the case of two electrons.
Given the difficulties in solving the general problem,
such analytical results (even for the most simple
configruations) are helpful. They are particularly useful for
nanostructures, where one might design controlled
experiments. \cite{zhitenev,tarucha}

We have recently reported on several exact results for two
interacting electrons on a general number of dots ${\cal N}$, which are
modeled
as `impurities' which have single electronic states:
we presented a general scheme for finding the eigenenergies,
and presented some results for the spectra of
a single dot. \cite{weprb}
We have also discussed the exact two-electron current
through a single dot. \cite{epl}
Here we generalize these results, with emphasis
on the charge accumulated on each quantum dot
and on its relationship with the Coulomb blockade picture.
We then devote most of this paper to discuss the more complex case of a
double dot. Such a double dot,
with one state per dot, has recently been proposed as a possible
candidate
for the two-qubit entanglement required for quantum computation.
\cite{das}  The case of two coupled quantum dots 
is also amenable to experiments. \cite{Fuji}

In our earlier work \cite{weprb}, we showed that
the spectrum and the wave functions of the two interacting
electrons can be obtained in terms of the energy
spectrum and the wave functions of the single-electron Hamiltonian.
We reproduce these results in section II in a slightly different
method,
and use them to obtain new results for
the {\it charge occupancies} on the quantum dots. The
next two sections are devoted to the study of specific configurations:
a single dot, and a system made up of two dots, separated
by a distance $R$. The single-electron spectra of these two
configurations, required for the study of the two electron one, are
discussed in the Appendix.

\section{Two interacting electrons -- general scheme}

As has been demonstrated in Ref. \onlinecite{weprb}, the knowledge of
the  spectrum of the single-electron Hamiltonian is sufficient for
deducing the spectrum and the wave functions of two interacting
electrons, for any number ${\cal N}$ of dots.
Basically, we start with the Hamiltonian
\begin{equation}
{\cal H}={\cal H}_{\rm se}+{\cal H}_{\rm int}.
\end{equation}
The spin-independent
single-electron part ${\cal H}_{\rm se}$ involves site energies
$\epsilon_{\rm i}$ on the dots (${\rm i}=1,~2,...,{\cal N}$) and
zero on the lead sites, and also nearest neighbor hopping matrix
elements
$t_{n,m}$ which assume special values near the dots.
This part is diagonalized by the eignenergies $\{\epsilon_a\}$
and the corresponding eigenfunctions $\{\phi_{a}(n)\}$.

For simplicity, we assume that the two electrons interact
only when they are both on the same dot i, with interaction energy
$U({\rm i})$ (though the method of
solution can be extended for other types of interactions):
\begin{equation}
{\cal H}_{\rm int}=\sum_{\rm i}U({\rm i})
c^{\dagger}_{{\rm i}\uparrow}c_{{\rm i}\uparrow}c^{\dagger}_{{\rm
i}\downarrow}
c_{{\rm i}\downarrow}.
\end{equation}
Using the single-electron eigenstates, the
two-electron Hamiltonian takes the form
\begin{eqnarray}
&&{\cal H}=\sum_{a\sigma}\epsilon_{a}c^{\dagger}_{a\sigma}c_{a\sigma}
+\sum_{{\rm i}}\sum_{abcd}U_{acbd}({\rm
i})c^{\dagger}_{a\uparrow}c_{b\uparrow}c^{\dagger}_{c\downarrow}
c_{d\downarrow},\nonumber\\
&&U_{acbd}({\rm i})=U({\rm i})\phi^{\ast}_{a}({\rm
i})\phi^{\ast}_{c}({\rm i})\phi_{b}({\rm i})\phi_{d}({\rm i}).\label{U}
\end{eqnarray}
Here, $c^{\dagger}_{a\sigma} \equiv
\sum_{n}\phi_{a}(n)c^{\dagger}_{n\sigma}$
creates an electron in the state
$a$ with spin $\sigma$.

For such a contact electron-electron interaction, one is interested
only in the {\it singlet}
 state of the two electrons (the energies of the
two electrons in the triplet state are simply given by the 
non-interacting sums
$\epsilon_a+\epsilon_b$).
We hence write for the two-electron singlet wave function
\begin{eqnarray}
|\Psi\rangle
=\sum_{ab}X_{ab}(E)c^{\dagger}_{a\uparrow}
c^{\dagger}_{b\downarrow}|0\rangle , \label{Psi}
\end{eqnarray}
where $|0\rangle $ is the vacuum and $X_{ab}=X_{ba}$.
The Schr\"{o}dinger equation
\begin{eqnarray}
{\cal H}|\Psi \rangle =E|\Psi \rangle
\end{eqnarray}
then yields
\begin{eqnarray}
&&\sum_{ab}\Bigl (E-\epsilon_{a}-\epsilon_{b}\Bigr
)X_{ab}(E)c^{\dagger}_{a\uparrow}c^{\dagger}_{b\downarrow}|0\rangle
\nonumber\\
&=&\sum_{{\rm i}}\sum_{aba' b'}X_{ab}(E)U_{a'b'ab}({\rm
i})c^{\dagger}_{a'\uparrow}c^{\dagger}_{b'\downarrow}|0\rangle .
\end{eqnarray}
Multiplying this equation from the left by $\langle
0|c_{b''\downarrow}c_{a''\uparrow}$ gives
\begin{eqnarray}
X_{ab}(E)=\sum_{{\rm i}}\sum_{a'b'}\frac{U_{aba'b'}({\rm
i})X_{a'b'}(E)}{E-\epsilon_{a}-\epsilon_{b}}.\label{X}
\end{eqnarray}
The simple form of the matrix elements of the contact
interaction  [see Eq.
(\ref{U})] allows us to rewrite Eq. (\ref{X}) as a set of ${\cal N}$
linear
equations:
Defining the quantities
\begin{eqnarray}
A_{{\rm i}}(E)=\sum_{ab}\phi_{a}({\rm i})\phi_{b}({\rm i})X_{ab}(E)
\label{A}
\end{eqnarray}
(which represent the amplitudes of $|\Psi \rangle$ for the singlet
state with both electrons on site i, denoted by $|{\rm i, i} \rangle$), 
one arrives at
\begin{eqnarray}
A_{{\rm i}}(E)=\sum_{{\rm i}'}U({\rm i}')G_{E}({\rm i,i;i',i'})A_{{\rm
i}'}(E),\label{det}
\end{eqnarray}
in which $G_{E}({\rm i,i;i',i'})$ 
is the two-particle Green's function of two {\it
non-interacting} electrons, 
\begin{eqnarray}
G_{E}({n_1,n_2;n_1',n_2'})=\sum_{ab}\frac{\phi_{a}(n_1)
\phi_{b}(n_2)\phi^{\ast}_{a}(n_1')\phi^{\ast}_{b}(n_2')
}{E-\epsilon_{a}-\epsilon_{b}},\label{GE}
\end{eqnarray}
calculated at the impurity locations. \cite{oppen}
The determinant of Eqs. (\ref{det}) gives the eigenenergies $\{E\}$ of
the two interacting electrons, and in particular determines the ground 
state energy, $E_G$. The coefficients $X(E)$ are then obtained
from Eq. (\ref{X}), which can be rewritten as
\begin{eqnarray}
X_{ab}(E)=\sum_{{\rm i}}\frac{U({\rm
i})\phi^{\ast}_{a}({\rm
i})\phi^{\ast}_{b}({\rm i})A_{\rm
i}(E)}{E-\epsilon_{a}-\epsilon_{b}}.\label{XX}
\end{eqnarray}
Substituting this result into Eq. (\ref{Psi}), it is easy to check that
\begin{equation}
|\Psi \rangle =\sum_{\rm i}U({\rm i})A_{\rm i}(E) \sum_{n_1,n_2}
G_E(n_1,n_2;{\rm i},{\rm i})
|n_1,n_2 \rangle,
\label{psig}
\end{equation}
$|n_1,n_2 \rangle$ is a singlet state
with one electron at site $n_1$ and the other at site $n_2$.
Note that Eq. (\ref{det}) determines the $A_{\rm i}$'s only up to a 
multiplicative
constant. This constant should be determined by the normalization of
$|\Psi \rangle$, i. e. from the condition $\sum_{ab}|X_{ab}(E)|^2=1$.
Such solutions will be discussed in some detail below.

The electronic states on a quantum dot are commonly probed by varying
the gate voltages on the dots, represented here by the $\epsilon_{\rm
i}$'s,
and measuring the conductance.
Alternatively, one may probe the total charge on the dots, by measuring
the
capacitance when the
gate voltage is changed.\cite{zhitenev} The total charge on the dots,
in the state $|\Psi \rangle$, is given by (in units of the electron
charge, $e$)
\begin{eqnarray}
P&=&\sum_{\sigma}\sum_{{\rm i}=1}^{{\cal
N}}\langle\Psi|c^{\dagger}_{{\rm i}\sigma}c_{{\rm
i}\sigma}|\Psi\rangle \nonumber\\
&&\equiv
\langle\Psi|\sum_{{\rm i}=1}^{{\cal N}}\sum_{ab}\phi^{\ast}_{a}({\rm
i})\phi_{b}({\rm
i})\sum_{\sigma}c^{\dagger}_{a\sigma}c_{b\sigma}|\Psi\rangle.
\end{eqnarray}
Using
Eq. (\ref{Psi}), we obtain
\begin{eqnarray}
P=2\sum_{{\rm i}}\sum_{abc}\phi^{\ast}_{a}({\rm i})\phi_{b}({\rm
i})X^{\ast}_{ac}(E)X_{bc}(E).\label{P}
\end{eqnarray}
Alternatively, we note that
\begin{equation}
P=\sum_{\rm i} \frac{\partial E}{\partial \epsilon_{\rm i}}.
\label{PP}
\end{equation}
This follows from first-order perturbation theory: writing $E=
\langle\Psi|{\cal H}|\Psi\rangle$, the derivative with respect to
$\epsilon_{\rm i}$ becomes
$\langle\Psi|c^{\dagger}_{{\rm i}\sigma}c_{{\rm
i}\sigma}|\Psi\rangle$.
In the following sections, we will use these results to discuss the
ground state of simple quantum dot systems occupied by two electrons.

An important technical advantage of the representation of the
two-electron
spectrum in terms of the two-particle Green's function of the
non-interacting system, is that the latter can be expressed in terms
of the {\it single-particle} Green's function. The spectral
representation of the single-particle Green's function, $g_{\omega}
(n;n')$, is
\begin{eqnarray}
g_{\omega}(n;n')=\sum_{a}\frac{\phi_{a}(n)\phi^{\ast}_{a}(n')}{\omega
+i\zeta -\epsilon_{a}},
\label{gg}
\end{eqnarray}
where $\zeta\rightarrow 0^{+}$. In conjunction with Eq. (\ref{GE}),
one finds \cite{epl}
\begin{eqnarray}
&&G_{E}(n_{1},n_{2};n_{1}',n_{2}')\nonumber\\
&=&-\frac{1}{\pi}\int_{-\infty}^{\infty}d\omega
g_{E-\omega}(n_{1};n_{1}')\Im g_{\omega}(n_{2};n_{2}')\nonumber\\
&=&\frac{i}{2\pi}\int_{-\infty}^{\infty}d\omega
g_{E-\omega}(n_{1};n_{1}')g_{\omega}(n_{2};n_{2}'),\label{Gg}
\end{eqnarray}
where the last equality follows from the Kramers-Kronig relations. The
relation (\ref{Gg}) is very useful in the detailed calculations of
the ground state properties, because the single-particle Green's
functions are relatively easy  to find. We give in Appendix A the
details of the single-particle Green's functions required for the
dot configurations investigated in this paper.

\section{a single dot on a one-dimensional wire}

We model a single dot on a one-dimensional wire by
a single impurity, located at site 0, which has the
on-site energy $\epsilon_{0}$, and is coupled to two ideal
one-dimensional leads, with the amplitude $t_{0}$ for tunneling
between the impurity and its nearest neighbors on the leads. \cite{ng}
Both these
parameters are experimentally accesssible: $\epsilon_{0}$ models
the plunger gate voltage on the dot,
while $t_{0}$ is related to the transmittance (or
conductance)
of the barriers between each  dot and the leads, which can be varied by
changing the gate voltages on these barriers. The
corresponding amplitudes between sites
inside the leads are equal to $-t$.
Although some aspects of the solution of this problem were discussed in
Ref.
\onlinecite{weprb}, we present here an alternative derivation, which
is more adapted to the calculation of the occupation $P$ and
to the more complicated case of two dots.

As shown in Appendix A,\cite{weprb} the single-particle Hamiltonian of
such a
system has none, one or two bound states, depending on the values of
the ``gate voltage" $\epsilon_{0}$ and the ``hybridization"
$t_{0}^{2}\equiv \gamma$
(energies are measured in units of $t$).
For the sake of
concreteness, we concentrate on the
region where there is only one bound state below the band,
of energy $\epsilon_{\beta}<-2$. This occurs for
$\epsilon_{0}<\epsilon_{00} \equiv 2(\gamma -1 )$ (see Appendix A).

In the simple case of a single `impurity',
Eq. (\ref{det}) reduces to a single equation, and
the eigenenergies $\{E\}$ of the two interacting electrons are given by
the solutions of
\begin{eqnarray}
\frac{1}{U}=G_{E}(0,0;0,0)=\sum_{ab}\frac{|\phi_{a}(0)|^{2}
|\phi_{b}(0)|^{2}}{E-\epsilon_{a}-\epsilon_{b}}.\label{G1U}
\end{eqnarray}
It is easy to deduce the beahvior of $G_{E}$ as function of the
two-electron energy $E$ from this equation. $G_{E}$ is negative for
$E<2\epsilon_{\beta}$, decreasing from 0 to $-\infty$ as $E$ increases
from $-\infty$ towards
$2\epsilon_{\beta}$.
As $E$ crosses this value, it jumps to $+\infty$ and then
decreases. The value $E=-2+\epsilon_{\beta}$ marks the beginning of the
two-particle continuous
band states: one electron is bound and the other is in the continuum.

\begin{figure}
\centerline{\psfig{figure=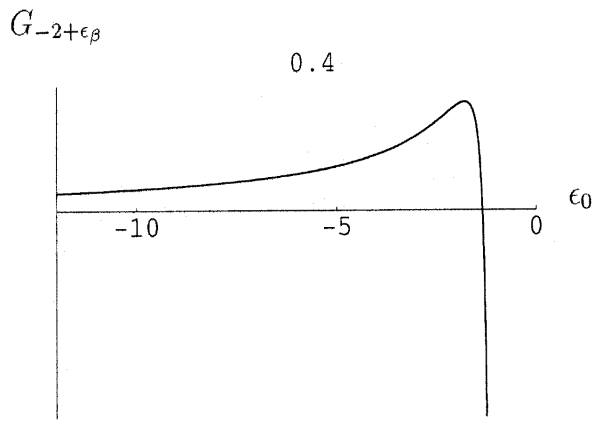,width=8cm,height=6cm}}
\vspace{.5cm}
\caption{$G_{E}(0,0;0,0)$ at $E=-2+\epsilon_{\beta}$,
as function of
$\epsilon_{0}$. The hybridization is fixed at $\gamma =0.4$.}
\label{Gthre}
\end{figure}

As discussed in Ref. \onlinecite{weprb},
$G_E$ is {\it finite} at $E=-2+\epsilon_{\beta}$ in the thermodynamic
limit
of infinite leads, due to the vanishing of
the band state wavefunction $\phi_k$ (with energy $\epsilon_k=-2\cos
k$)
at the impurity site ${\rm i}=0$ for $k=0$.
This value of $G$ determines whether
there is or there is not a bound state of the {\it two interacting
electrons}: When $1/U<G_{-2+\epsilon_{\beta}}$, then Eq. (\ref{G1U})
has no solution for $E<-2+\epsilon_\beta$, and there is
no doubly occupied bound state below the band.
One of the electrons is then in a band state.
The
behavior of $G$ at $E=-2+\epsilon_{\beta}$, as
function of $\epsilon_{0}$, is plotted in Fig. \ref{Gthre}. Since
$G_{E=-2+\epsilon_{\beta}}$ has a maximum, $G_{\rm max}$, there is
always
a doubly occupied bound state (or an ``insulator") for $U<1/G_{\rm
max}$.
For larger $U$, the equation $G_{-2+\epsilon_\beta}=1/U$
has {\it two} solutions, $\epsilon_{0,-}$
and $\epsilon_{0,+}$ (which depend on $U$).
For $\epsilon_{0,-}<\epsilon_{0}<\epsilon_{0,+}$
one has no doubly occupied bound state, and
the ground state of the two electrons
lies in the continuum (i. e. represents a ``metal").
Then, as the on-site energy $\epsilon_{0}$
becomes more attractive, the bound state of the two electrons
re-appears.
As seen from Fig. \ref{Gthre}, $\epsilon_{0,-}$ diverges to $-\infty$
when $U \rightarrow \infty$, and this re-entrance then disappears.
The region between $\epsilon_{0,-}$ and $\epsilon_{0,+}$ becomes
narrower
as $\gamma$ increases, and for finite $U$ it always disappears above
some
critical hybridization $\gamma_c$
(which diverges to $\infty$ as $U \rightarrow \infty$).
\cite{weprb}

These ``insulator to metal"
\cite{foot}
transitions of the two-electron ground state, from being bound
to being in the continuum and back, are reflected in the occupancy $P$
of the
dot in the ground state [see Eq. (\ref{PP})].
To find $P$, we
re-write the equation for the ground energy $E_{G}$ in the form
\begin{equation}
\frac{1}{U}=-\frac{1}{\pi}\int_{-\infty}^{\infty}d\omega
(D_{E_{G}-\omega})^{-1}\Im (D_{\omega})^{-1},
\label{eg}
\end{equation}
where we have used Eq. (\ref{Gg}), and the results for $g_{\omega}$
relevant for this geometry [Eq. (\ref{D})].
The imaginary part appearing
in this expression has a delta-function contribution coming from the
single-electron bound energy, and the contribution arising from the
band states. Separating these two, we have
\begin{eqnarray}
\frac{1}{U}&=&\frac{r(\epsilon_{\beta})}{D_{E_{G}-\epsilon_{\beta}}}\nonumber\\
&-&\frac{1}{\pi}\int_{-2}^{2}d\omega
(D_{E_{G}-\omega})^{-1}\Im (D_{\omega})^{-1},\label{FGU}
\end{eqnarray}
where $r(\omega )=(\partial D_{\omega }/\partial \omega )^{-1}$ is the
residue at the bound energy pole $\omega=\epsilon_\beta$.
It is now straightforward to
differentiate all terms in (\ref{FGU}) with respect to $\epsilon_{0}$,
and obtain $P=\partial E_{G}/\partial\epsilon_{0}$.

\begin{figure}
\centerline{\psfig{figure=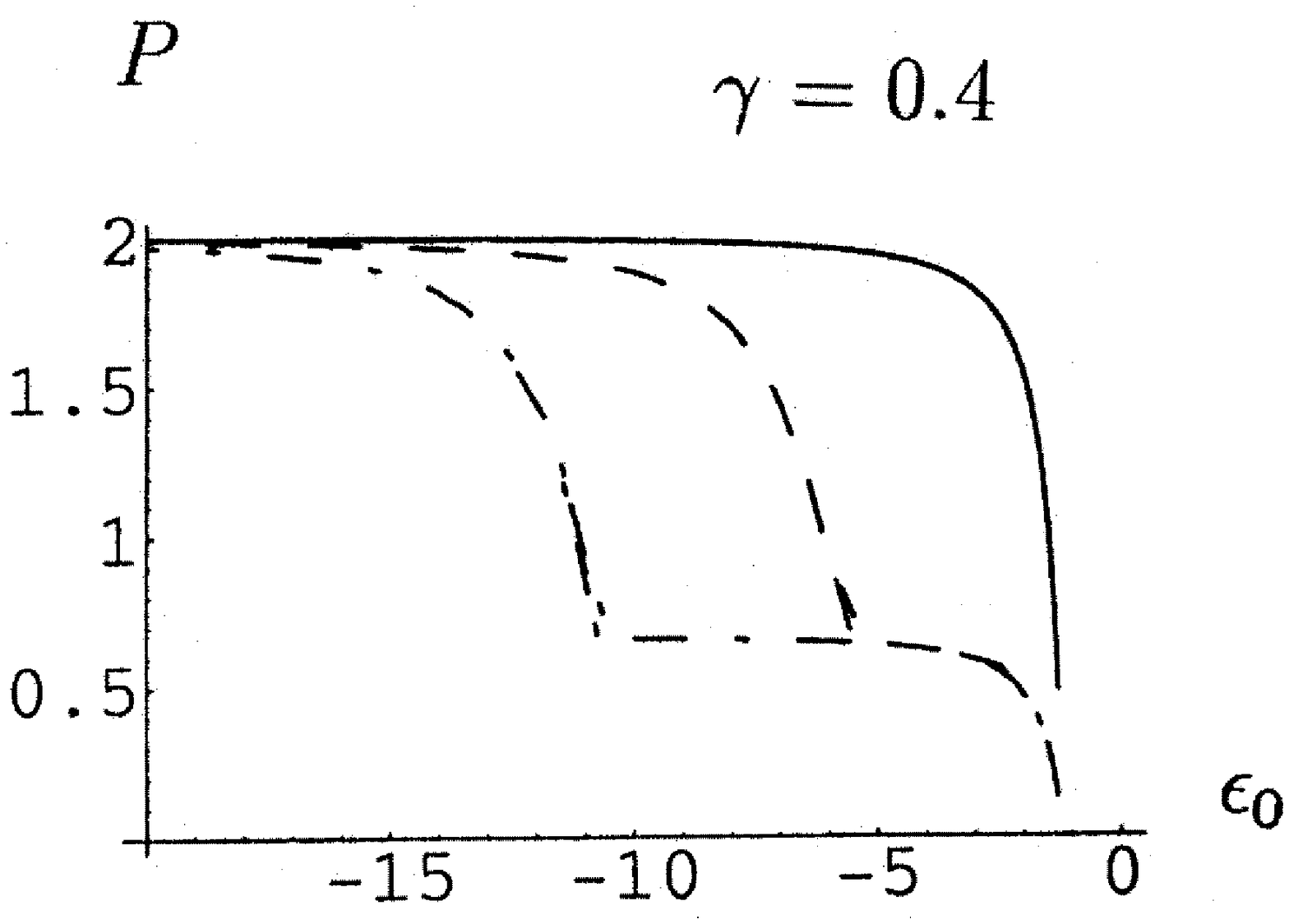,width=8cm,height=6cm}}
\vspace{2cm}
\caption{The occupancy of the single dot, as function of
$\epsilon_{0}$. The curves, from right to left, are for $U$=2, $U=5$,
and $U=10$, respectively. The hybridization is fixed at $\gamma =0.4$.}
\label{Pg4}
\end{figure}

\begin{figure}
\centerline{\psfig{figure=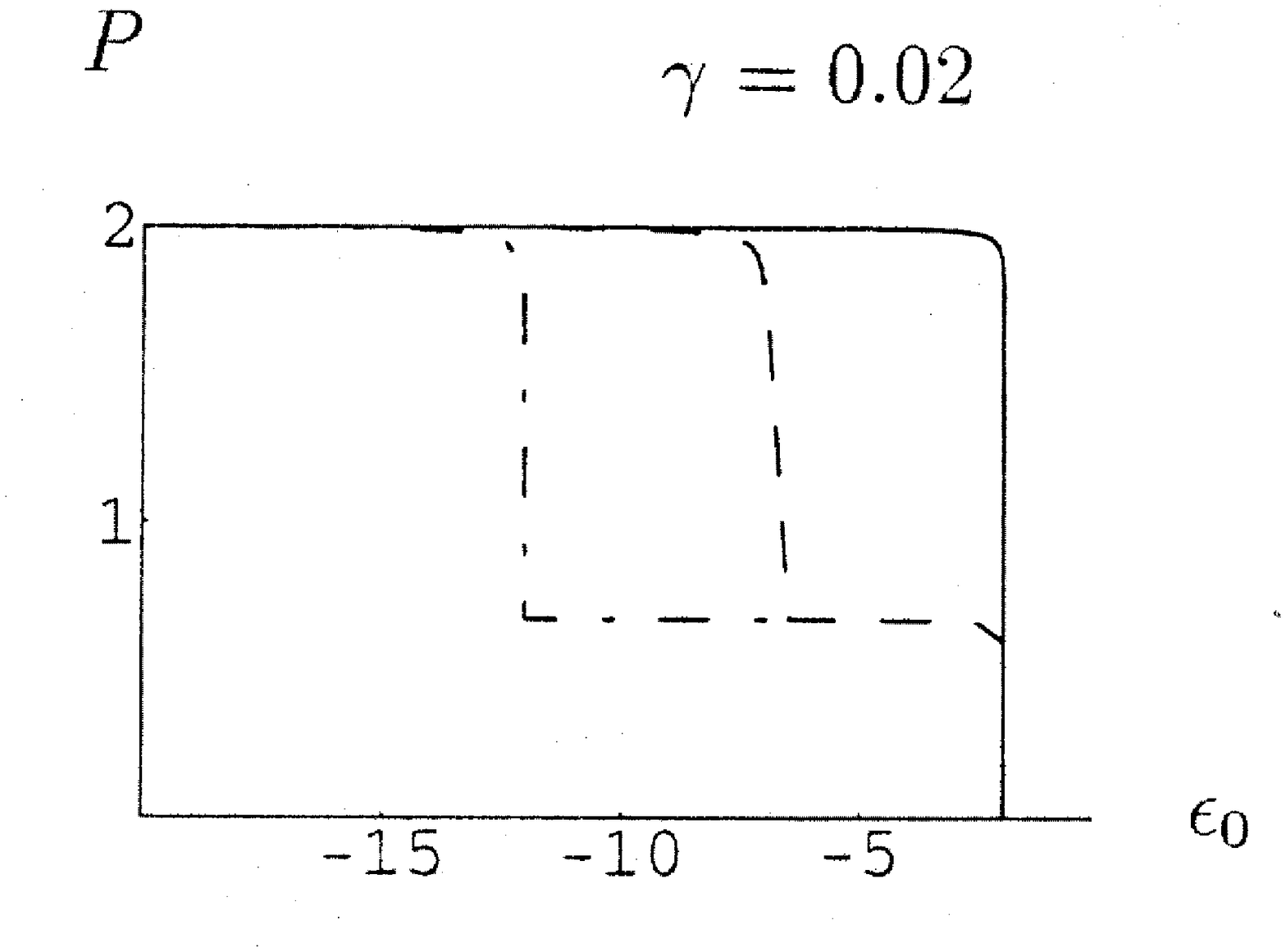,width=8cm,height=6cm}}
\vspace{2cm}
\caption{The occupancy of the single dot, as function of
$\epsilon_{0}$. The curves, from right to left, are for $U$=2, $U=5$,
and $U=10$, respectively. The hybridization is fixed at $\gamma =0.02$.}
\label{Pg05}
\end{figure}

We have solved Eq. (\ref{FGU}) for $E_G$, by calculating the integral
numerically. We have then
computed the variation of the occupancy $P$ as function of the gate
voltage $\epsilon_{0}$, and the results are shown in Fig. \ref{Pg4},
for a comparatively large hybridization, and in Fig. \ref{Pg05}, for a
small hybridization $\gamma $.
Generally, $P$ starts at 0 at $\epsilon_0=\epsilon_{00}=2(\gamma-1)$, 
when the single
electron bound state just moves below the band (with
the inverse localization length $\kappa_\beta=0$, i. e.
zero weight on the impurity). $P$ then grows as $\epsilon_0$ decreases.
As seen from Fig. \ref{Gthre}, $\epsilon_{0,+}$ is quite close to
$\epsilon_{00}=2(\gamma-1)$, 
due to the steepness of $G$ [$G$ diverges to $-\infty$
as $\epsilon_0 \uparrow 2(\gamma-1)$]. Therefore, the first
transition into the ``metallic" phase, at $\epsilon_{0,+}$,
occurs when the localization lengths
of the two bound electrons are still quite large, and their weights on
the impurity (and thus also $P$)
are relatively small. The calculation of $P$ in this regime is
not easy, due to numerical problems related to the above mentioned
steepness. In any case, $P$ reaches values close to 1
somewhere inside the ``metallic"
phase, i. e. for $\epsilon_{0,-}<\epsilon_0<\epsilon_{0,+}$.
We note that in the first ``insulating" phase, which appears at 
$\epsilon_{0,+}<\epsilon_0<2(\gamma-1)$, both electrons are bound
on very shallow states, hence the small value of $P$. Thus, it is not
enough to know $P$ in order to determine the transport nature of the system. 
As $\epsilon_0$ crosses below
$\epsilon_{0,-}$, into the second ``insulating" phase,
$P$ gradually increases towards
2, reflecting the strongly bound state of the two electrons.
This gradual increase becomes steeper as the hybridization $\gamma$
becomes smaller, and the width of the ``metallic"
single electron occupancy regime
(of order $\epsilon_{0,+}-\epsilon_{0,-}$)
increases with increasing $U$. Both of these facts are in {\it
qualitative}
accordance with
the Coulomb blockade picture (where usually the {\it derivative} of $P$
with respect to the gate voltage has peaks whose width increases with
the hybridization and whose inter-peak distance increases with $U$).
\cite{CB,M}
In fact, the distance between the $N$'th and the $(N-1)$'th
peaks
is usually interpreted as the energy cost of adding the $N$'th electron.
However, it is usually very difficult to obtain quantitative estimates
for these quantities in that picture. Furthermore, the similarity of 
our results to the simple Coulomb blockade picture is completely lost
as $\gamma$
increases towards and beyond $\gamma_c$: the width of the ``metallic" regime
then shrinks, and the there is a continuous gradual increase of $P$
from 0 to 2.

Returning to Eq. (\ref{psig}), we now observe that for a single impurity,
the two-electron state is given by 
\begin{equation}
|\Psi \rangle = UA\sum_{n_1,n_2}G_E(n_1,n_2;0,0)|n_1,n_2 \rangle,
\end{equation}
where $A$ is found from the normalization.
One can now use Eq. (\ref{Gg}) and the single electron Green's functions
$g_\omega(n;0)$ to obtain $|\Psi \rangle$. For the bound ground state,
the results show an exponential decay of the amplitudes as either electron
moves away from the impurity.


\section{Two dots on a one-dimensional wire}

Two dots are modeled by two `impurities', connected to each other and to
the
outside
by ideal linear leads.
The presence of two impurities gives rise to up to four single-particle
bound states. For simplicity, we consider two identical impurities,
each having the same on-site energy $\epsilon_{0}$, which are located
at sites $\ell $ and $r$, and are separated by a distance $R$ ($R \ge
2$).
Confining ourselves again to the configuration where the bound
states appear
only below the band, the first bound state appears when
$\epsilon_{0}$ is
smaller than $\epsilon_{00}=2(\gamma -1)$, while the second appears only for
$R>R_{c}$, where
\begin{eqnarray}
R_{c}=2\gamma /(2\gamma -2 -\epsilon_{0}).\label{RC}
\end{eqnarray}
At fixed $R$, there exists a single bound state below the band only
in the narrow regime
\begin{equation}
2(\gamma-1-\gamma/R)<\epsilon_0<2(\gamma-1).
\label{epspm}
\end{equation}
We also restrict ourselves to the regime with $\epsilon_0<2(1-\gamma)$,
so that there are no bound states above the band (see Appendix A).

Assuming the on-site Coulomb interaction to be identical
on the two impurities, $U(\ell)=U(r)\equiv U$, Eqs. (\ref{det}) yield
\begin{eqnarray}
A_{\ell }(E)&=&UG_{E}({\rm d} )A_{\ell
}(E)+UG_{E}({\rm nd})A_{r}(E),\nonumber\\
A_{r }(E)&=&UG_{E}({\rm nd} )
A_{\ell }(E)+UG_{E}({\rm d})A_{r}(E),
\end{eqnarray}
where the labels d and nd stand for the diagonal and the nondiagonal
elements
of the matrix. By symmetry,
\begin{eqnarray}
G_{E}({\rm d})&\equiv &G_{E}(\ell ,\ell ;\ell ,\ell
)=G_{E}(r,r;r,r),\nonumber\\
G_{E}({\rm nd})&\equiv &G_{E}(\ell ,\ell ;r,r)=G_{E}(r,r;\ell ,\ell ).
\end{eqnarray}
The eigenenergies of
the two interacting electrons are given by the solutions of the two
equations
\begin{eqnarray}
\frac{1}{U}=G_{E}({\rm d} )\pm G_{E}({\rm nd})\equiv
G^{\pm}_{E},\label{eigen}
\end{eqnarray}
and the corresponding solutions obey
\begin{equation}
A^\pm_\ell(E)=\pm A^\pm_r(E).
\label{Apm}
\end{equation}

It is instructive to rewrite these equations in terms of the
single-electron
wave functions, using Eq. (\ref{GE}).
In the symmetric molecule case, one can divide the solutions into 
even and odd
single-electron wave functions, with $\phi_a(\ell)=\pm \phi_a(r)$.
From Eq. (\ref{GE}) it now follows that
\begin{eqnarray}
G_E^\pm=\sum_{ab}\frac{[\phi_{a}(\ell)
\phi_b(\ell) \pm \phi_a(r)\phi_b(r)]\phi^\ast_a(\ell)\phi^\ast_b(\ell)}
{E-\epsilon_{a}-\epsilon_{b}}.
\end{eqnarray}
Thus, it is clear that $G_E^+$ contains only pairs of states where both
$a$ and $b$ are even or odd, while $G_E^-$ contains only mixed
combinations,
where one state is even and the other is odd.
It thus follows that the solutions of the two-electron problem divide
into two separate families: the solutions of $G_E^+=1/U$ involve only
even-even and odd-odd single electron states, while those of 
$G_E^-=1/U$
involve only even-odd states: the coefficients in Eq. (\ref{Psi})
will split into two separate families, associated with the different
solutions of Eqs. (\ref{eigen}). This can be easily seen by substituting
Eq. (\ref{Apm}) into Eq. (\ref{psig}).

To discuss the two-electron energies, we need to analyze the
$E$-dependence
of $G_E^\pm$.
This depends on $R$: For $R>R_c$, there exist two single-electron bound
states,
the even $\phi_{\beta +}$ and the odd $\phi_{\beta -}$.
Thus, $G_E^+$ decreases from $\infty$
to $-\infty$ as $E$ increases from $2\epsilon_{\beta +}$ towards
$2\epsilon_{\beta -}$. Therefore, in this case
the equation $G_E^+=1/U$ always has a
discrete solution, with $E$ between $2\epsilon_{\beta +}$ and
$2\epsilon_{\beta -}$. In the same case, $G_E^-$ 
decreases from $\infty$
towards a finite value as $E$ increases from $\epsilon_{\beta
+}+\epsilon_{\beta -}$ towards the bottom of the continuum
$-2+\epsilon_{\beta +}$ (which
contains both even and odd states). The new lowest even-odd state may
thus be
either ``insulating" or ``metallic", depending on the sign of
$1/U-G^-_{-2+\epsilon_{\beta +}}$. However, the energy of this even-odd
state
is always above the lowest triplet energy, which is equal to
the non-interacting value $\epsilon_{\beta +}+\epsilon_{\beta -}$.
From our numerical calculations we observe that $G_E^+$ is negative at
this lowest triplet energy. Therefore, the
lowest solution of $G_E^+=1/U$ is the ground state of the two-electron
problem, which is thus a singlet.
In a way, this might have been expected: breaking the system into two
parts, by removing a bond in the middle between the two dots, one
ends up with separate ``atomic" states on each side,
each coupled to its own lead. Each side can then
contain one electron with either spin up or spin down. However,
switching on
the hopping $t_h$ between the two sides would lower the energy of the
singlet
state, similarly to the antiferromagnetic ground state of the Hubbard
model;
to lowest order in $t_h$, the ``exchange" difference between the
triplet and singlet states is of order $t_h^2/U$.\cite{exchange}
It is interesting to note that in our case there exists a finite
difference
between the singlet and the triplet even in the limit $U \rightarrow
\infty$,
since we find that $G^+_{\epsilon_{\beta +}+\epsilon_{\beta -}}$ is
strictly
negative.
It would be interesting to study generalizations of our simple model,
e. g. including interdot Coulomb and exchange interactions,
which would allow an interchange of the singlet and triplet ground
states.
\cite{tarucha1}

The only chance to find a ``metallic" ground state is thus for $R<R_c$,
when there exists only one single-electron bound state below the band.
This limits the possible range of parameters to
that in Eq. (\ref{epspm}).
In this regime, $G_E^-$ yields no doubly bound state, and $G_E^+$ yields
one
only if $G^+_{-2+\epsilon_{\beta +}}<1/U$.
Note that this regime becomes narrower (in terms of $\epsilon_0$)
as $R$ increases.
It is therefore interesting to find the borderline in the
parameter space, at which $G^{+}_{-2+\epsilon_{\beta +}}=1/U$.
Inside this broderline,
the ground energy of the two electrons is in the continuum, i. e.
``metallic".

To calculate $G_E^\pm$, we
use Eq. (\ref{Gg}) and the results (\ref{gdn}) and
(\ref{DPM})   of Appendix A:
\begin{eqnarray}
&&G^{+}_{E}=G_{E}({\rm d})+G_{E}({\rm nd})\nonumber\\
&=&-\frac{1}{2\pi}\int d\omega \Biggl (
\frac{1}{D_{E-\omega}^{-}}\Im\frac{1}{D^{-}_{\omega}}+
\frac{1}{D_{E-\omega}^{+}}\Im\frac{1}{D^{+}_{\omega}}\Biggr
),\nonumber\\
&&G^{-}_{E}=G_{E}({\rm d})-G_{E}({\rm nd})\nonumber\\
&=&-\frac{1}{2\pi}\int d\omega \Biggl (
\frac{1}{D_{E-\omega}^{-}}\Im\frac{1}{D^{+}_{\omega}}+
\frac{1}{D_{E-\omega}^{+}}\Im\frac{1}{D^{-}_{\omega}}\Biggr ),
\end{eqnarray}
where $D^{\mp}_{\omega}$ are given by Eqs. (\ref{DPM}).

We next separate the contributions of the bound energies from the
integrals,
to find
\begin{eqnarray}
&&G^{+}_{E}=\frac{1}{2}\Biggl
(\frac{r^{+}(\epsilon_{\beta +})}{D^{+}_{E-\epsilon_{\beta +}}}+\Theta
(R
-R_{c})\frac{r^{-}(\epsilon_{\beta -})}{D^{-}_{E-\epsilon_{\beta
-}}}\Biggr
)\nonumber\\
&&-\frac{1}{2\pi}\int_{-2}^{2} d\omega \Biggl (
\frac{1}{D_{E-\omega}^{-}}\Im\frac{1}{D^{-}_{\omega}}+
\frac{1}{D_{E-\omega}^{+}}\Im\frac{1}{D^{+}_{\omega}}\Biggr
),\nonumber\\
&&G^{-}_{E}=\frac{1}{2}\Biggl
(\frac{r^{+}(\epsilon_{\beta +})}{D^{-}_{E-\epsilon_{\beta +}}}+\Theta
(R
-R_{c})\frac{r^{-}(\epsilon_{\beta -})}{D^{+}_{E-\epsilon_{\beta
-}}}\Biggr
)\nonumber\\
&&-\frac{1}{2\pi}\int_{-2}^{2} d\omega \Biggl (
\frac{1}{D_{E-\omega}^{-}}\Im\frac{1}{D^{+}_{\omega}}+
\frac{1}{D_{E-\omega}^{+}}\Im\frac{1}{D^{-}_{\omega}}\Biggr
),\label{workG}
\end{eqnarray}
where $r^{\pm}(\omega )=(\partial D^{\pm}_{\omega}/\partial\omega
)^{-1} $ are
the residues at the poles.

\begin{figure}
\centerline{\psfig{figure=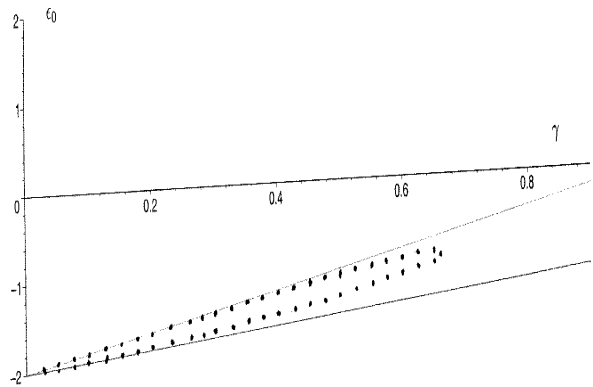,width=8cm,height=6cm}}
\vspace{2cm}
\caption {$\epsilon_0$--$\gamma$ phase diagram for $R=2$. The straight
lines
which meet at $\epsilon_0=-2$ represent the bounds in Eq. (\ref{epspm}),
between which there exists only one single-electron bound state below
the band.
Inside the dotted curved
line, the system is ``metallic", for $U=\infty$.}
\label{phd}
\end{figure}

We have used  Eq. (\ref{workG}) to solve the equation
 $G^+_{-2+\epsilon_{\beta }}=0$,
which yields the borderline of the ``metallic" regime in the limit $U
\rightarrow
\infty$. The result for $R=2$
is depicted by the dotted line in Fig. \ref{phd}.
The area enclosed inside this line represents the ``metal", where
$G^{+}_{-2+\epsilon_{\beta +}}>0$.
For smaller $U$ and for
larger $R$'s this area shrinks
further.

Figure \ref{phd} highlights a major difference between the single-dot
and the double-dot cases. In the former, the width of the 
``metallic" regime
(in terms of $\epsilon_0$) was equal to 
$\epsilon_{0,+}-\epsilon_{0,-}$,
and at fixed $\gamma$ it increased with $U$, diverging to $\infty$ for
$U \rightarrow \infty$. Although this width was not equal to $U$, as
assumed in the simple Coulomb blockade picture, it still resembled
the qualitative features of that picture.
In contrast, in the double-dot case this width is bounded by Eq.
(\ref{epspm}),
and this bound is {\it independent} of $U$. Therefore, the width
of the singly occupied ``metallic" regime remains finite
and small even when $U \rightarrow \infty$.
Basically, this happens because in
the double-dot case, there exist {\it two}
single-electron
bound states.
The level-repulsion between these bound states then
prevents the two-electron ground state
from
merging into the coninuum. In fact, we expect similar bounds on the 
Coulomb-blockade-like energy even for a single quantum dot,
whenever the dot has more than
a single bound state.

\begin{figure}
\centerline{\psfig{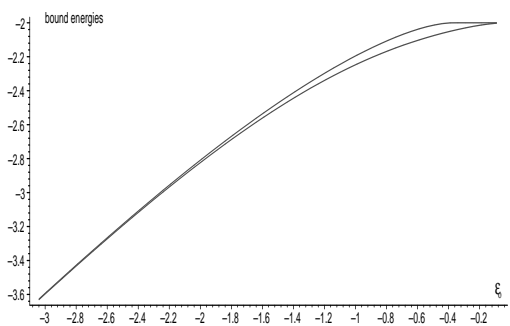}}
\vspace{-5cm}
\caption{$\epsilon_{\beta +}$ and $\epsilon_{\beta -}$ as function of
$\epsilon_{0}$. Here $R=6$, and the dot is `open': $\gamma =0.98$. }
\label{ben}
\end{figure}

\begin{figure}
\centerline{\psfig{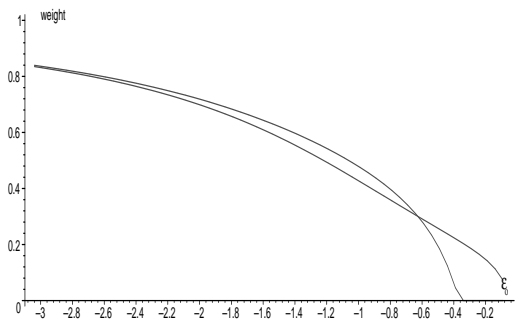}}
\vspace{-5cm}
\caption{The weight of the single-particle wave
functions in the bound states on the impurities.
Here $R=6$, and the dot is `open`: $\gamma =0.98$. }
\label{weight}
\end{figure}

As $R$ increases,
the two single-electron ground energies of the double dot become closer
to each other and to the single-dot bound state energy. Since
$2\epsilon_{\beta +}<E_G<2\epsilon_{\beta -}$, it follows that
the ``insulating" ground energy of the two interacting
electrons is almost independent of $U$.
Moreover, as $|\epsilon_{0}|$
increases, the two single-particle bound energies approach one
another, and practically we have only one,
doubly-degenerate,
single-particle bound energy. This behavior is shown in Fig.
\ref{ben}, for a very `open' dot. Similar effects arise
when the hybridization is reduced:
 $\epsilon_{\beta +}$ and $\epsilon_{\beta -}$ also become
indistinguishable. It
hence follows that independently of $U$, the ground state energy
of the two
interacting electrons is $E_{G}\simeq 2\epsilon_{\beta +}\simeq
2\epsilon_{\beta -}$. In such a situation, the charge 
accumulated on the
quantum dot, $P$, will just follow the weight of the single-particle
localized wave functions on the impurities. These have a `smooth'
behavior as function of $\epsilon_{0}$ (see Fig. \ref{weight}).
Consequently, the ``Coulomb blockade"
type behavior, which is obtained for
the single-impurity dot (Figs. \ref{Pg4} and \ref{Pg05}) is washed out.

Finally, we comment on the two-electron wave function in the ground state.
At small $R$, when there exists only one single-electron bound state,
this wave function is dominated by the even-even state in which both
electrons occupy the state $\phi_{\beta +}$. To leading order in $\gamma$
this term represents the Heitler-London molecular state, 
proportional to 
$|\ell,\ell \rangle+|r,r \rangle+|\ell,r \rangle +|r,\ell \rangle$, 
with equal weights to
the electrons being on the same impurity or each electron being on a different
impurity.  For large $R$, $E_G$ is close to both $2\epsilon_{\beta +}$
and to $2\epsilon_{\beta -}$, and therefore the coefficients $X_{ab}(E_G)$
will be dominated by the terms with $a,b = \beta +,\beta +$ or
$a,b = \beta +, \beta -$, with roughly equal magnitudes for these two
terms [see Eq. (\ref{XX})]. 
Combining these two terms, it is easy to see that $|\Psi \rangle$
is then dominated by the {\it atomic} orbitals $|\ell,r \rangle +
|r, \ell \rangle$. Thus, our exact results interpolate nicely between these
two leading approximations, which are common in chemistry textbooks.
\cite{harrison}

\section{conclusions}

Our exact solution of the two-electron problem does not have
the Fermi
gases on the leads; it is limited to the `canonical ensemble',
with a fixed small number of electrons in the system.
We do believe however that
this exact solution for
this simple case is still useful, 
in that it throws light on issues which
sometimes
remain unclear in approximate (and much more complicated)
treatments of the real
problem. 

Our solution may also be directly connected with
e. g. the ionization 
of donors into
the conduction band, as function of the system parameters.
Simple models for this problem may involve ${\cal N}$
coupled one electron donors, or a single
dielectronic donor. Our calculation shows that as the distance between 
a pair of single-electron donors decreases, then the
on-site interaction $U$ helps this ionization.
This effect  may well
be an important ingredient for the real metal-insulator transition in
some 
semiconductors. Specifically, if each donor in the semiconductor has
exactly one electron attached to it, then as the density
of donors increases we expect pair of donors to combine into ``molecules"
which allow only one bound electron. The remaining electrons will move to 
the band, and the system will become ``metallic".

The fact that even an onsite $U$ can have effects which are
so different from
the naive
Coulomb blockade model, should also be of interest. 
This is especially so for
the double quatum dot case.
As mentioned in the introduction, 
the spectra
of such systems are in principle addressable by transport and capacitance
experiments.
Usually, one looks at the contribution of the
resonant states (lying in the continuum) of these dots to the
conductivity,
but the bound states will also contribute to the off-resonance
transmission.
The dependence of the average occupancy on the gate voltage is of
interest
both theoretically \cite{tarucha} and experimentally.

Generalizations of this treatment to more realistic situations, even for
two electrons, are relatively easy to achieve. For example, the inclusion
of an interdot interaction $V$ does not affect the need to solve only
${\cal N}$ linear equations.
As stated, such interactions may cause an interchange of the singlet
and triplet ground states \cite{tarucha}, and will certainly
affect the
magnetic exchange
interactions between the electrons on different `impurities'.


\acknowledgements
This paper is dedicated to Franz Wegner, on the occasion of his 60th
birthday.
All authors acknowledge the hospitality of the Centre for Advanced
Studies
of the Norwegian Academy, where parts of this work were done.
This research is also supported by grants from the Israeli Science
Foundation and from the Israeli Ministry of Science and the French
Ministry of Research and Technology.

\appendix
\section{The single-particle Green's function}

As discussed in the text, the two-particle Green's function can be
expressed in terms of the single-particle one. Here we derive the
latter.

The single-electron tight-binding type Hamiltonian,
${\cal H}_{\rm se}$, is given by
\begin{eqnarray}
{\cal H}_{\rm
se}&=&\sum_{n}\epsilon_{n}c^{\dagger}_{n}c_{n} \nonumber\\
&+&\sum_{n}(t_{n,n+1}
c^{\dagger}_{n}c_{n+1}+t_{n-1,n}c^{\dagger}_{n}c_{n-1}),
\end{eqnarray}
where we ignore the spin indices, since the single particle Hamiltonian
is spin-independent.
Writing the single-particle Green's function, $g$, in the form
\begin{eqnarray}
g(n,n';t)=-i\Theta (t)\langle [c^{\dagger}_{n}(t),c_{n'}]_{+}\rangle ,
\end{eqnarray}
we find, for the Fourier transform $g_{\omega}$, the equation
\begin{eqnarray}
&&\omega g_{\omega}(n,n')=
\delta_{n,n'}+\epsilon_{n}g_{\omega}(n ,
n')\nonumber\\
&+&t_{n,n+1}g_{\omega}(n+1,n')+t_{n-1,n}g_{\omega}(n-1,n').\label{geom}
\end{eqnarray}
This equation is straightforwardly solved for the configurations
described in the text.

\subsection{The single impurity case}

For a single impurity with one-dimensional leads
we have $\epsilon_{n}=0$, for $n\neq 0$, and
$\epsilon_{0}$ is the on-site energy of the impurity,
$t_{n,n\pm 1}\equiv
-t$ for $n\neq 0, \pm 1$, and $t_{0,\pm 1}\equiv t_{0}$.
It is sufficient to consider Eq.
(\ref{geom})  for $n' =0$.
Then for any $n\neq 0$ or $\pm 1$ that equation gives
\begin{eqnarray}
\omega g_{\omega}(\pm n,0)=-tg_{\omega}(\pm (n+1),0)-tg_{\omega}(\pm
(n-1),0).
\label{A4}
\end{eqnarray}
It is easy to convince oneself that, in the limit of infinite leads,
the Green's function does not depend on the details of the boundary
conditions.
For $n \ne 0$ one can therefore assume the solution
\begin{equation}
 g_{\omega}(
n,0)=C_{\omega}a^{|n|}_{\omega},
\end{equation}
and find from Eq. (\ref{A4}) that
\begin{eqnarray}
\frac{\omega}{t}&=&-a_{\omega}-1/a_{\omega}.
\label{a}
\end{eqnarray}
Thus, for $|\omega/t|<2$ we can denote $a_\omega=e^{ik_\omega}$, with
$\omega/t=-2 \cos k_\omega$. For
$\omega /t){\stackrel{>}{<}}\pm 2$ we denote $a_\omega= \pm
e^{-\kappa_\omega}$,
and
$\omega/t=\mp 2\cosh \kappa_\omega$.

The equation for $n=\pm
1$ now yields $C_{\omega}=-(t_{0}/t)g_{\omega}(0,0)$, and finally the
equation for $n=0$ yields
\begin{eqnarray}
g_{\omega}(0,0)&=&1/D_{\omega},\nonumber\\
D_{\omega}&=&\omega -\epsilon_{0}+2(t_{0}^{2}/t)a_{\omega}\label{D}.
\end{eqnarray}

From Eq. (\ref{gg}), the poles of $g_\omega(0,0)$ give the eigenvalues
of the single electron problem, while the corresponding residues give
the
probability that an electron in a given state in on the impurity.
The boundary between having or not having a bound state is easily
found by setting $a_\omega=\pm 1$ and $\omega/t=\mp 2$
in the equation $D_\omega=0$ (with the upper sign refering to a state
below the band).
Measuring energies in units of $t$, and denoting
\begin{eqnarray}
\gamma =(t_{0}/t)^{2},
\end{eqnarray}
we find
that for $\epsilon_{0}<2(\gamma -1)$ there exists a bound
state below the band (with $0<a_\omega=e^{-\kappa_\beta} \le 1$),
with a localization length $1/\kappa_{\beta}$ [which diverges to
$\infty$
at $\epsilon_{0}=2(\gamma -1)$] and
energy $\epsilon_{\beta}=-2{\rm cosh}(\kappa_{\beta})$. For
$\epsilon_{0}>2(1 -\gamma )$ there appears a bound state above the
band, with a localization length $1/\kappa_{\alpha}$ and energy
$\epsilon_{\alpha}=2{\rm cosh}(\kappa_{\alpha})$. The two localization
lengths
are given by
\begin{eqnarray}
e^{\kappa_{\alpha,\beta}}=\pm\frac{\epsilon_{0}}{2}+\sqrt{\Bigl
(\frac{\epsilon_{0}}{2}\Bigr )^{2}-1 + 2\gamma}.
\end{eqnarray}
The weights of the localized wave function on the impurity (i.e., the
residues of $g_{\omega}$ at the bound energies) are accordingly
\begin{eqnarray}
|\phi_{\alpha,\beta}(0)|^{2}\equiv \frac{\partial
\epsilon_{\alpha,\beta}}
{\partial
\epsilon_{0}} =\Bigl
[1+\frac{2\gamma}{e^{2\kappa_{\alpha,\beta}}-1}\Bigr
]^{-1}.
\end{eqnarray}
A similar analysis gives the band of extended states, with energies
$\epsilon_k
=-2\cos k$.

\subsection{The two impurity case}

Here we consider a system with two impurities, which are separated by
a distance
$R$ ($R \ge 2$). We denote the locations of the two impurities by
$\ell $ and $ r$, and assume $t_{n,n\pm 1}\equiv -t$ for $n\neq\ell$
or $r$, $t_{n,n\pm 1}\equiv t_{0}$ for $n=\ell$ or $r$,
$\epsilon_{n}=0$ for $n\neq\ell$ or $r$, and $\epsilon_{\ell
,r}\equiv\epsilon_{0}$.

Again, it is sufficient to consider
$g_{\omega}(n,n')$ with $n'=\ell,~r$.
Referring to Eq. (\ref{geom}),
we assume a solution of the form
\begin{eqnarray}
g_{\omega}(n,\ell )&=&C^{<}_{\omega}(\ell )a^{|n-\ell |}_{\omega},
\ \ \ n<\ell ,\nonumber\\
g_{\omega}(n,\ell )&=&C^{>}_{\omega}(\ell )a^{|n-r|}_{\omega},
\ \ \ n>r,\nonumber\\
g_{\omega}(n,\ell )&=&A_{\omega}(\ell )a^{n}_{\omega}+B_{\omega}(\ell
)a^{-n}_{\omega},\ \ \ell <n<r,
\end{eqnarray}
where $a_{\omega}$ is the solution of Eq. (\ref{a}). Writing Eq.
(\ref{geom})
for $n= \ell -1$ and $n=r+1$ gives
\begin{eqnarray}
C^{<}_{\omega}(\ell )&=&-\frac{t_{0}}{t}g_{\omega}(\ell ,\ell), \ \
C^{>}_{\omega}(\ell )=-\frac{t_{0}}{t}g_{\omega}(r,\ell ).
\end{eqnarray}
The other two coefficients, $A_{\omega}(\ell )$ and $B_{\omega}(\ell
)$ are found by using the equation for $n=\ell +1$ and $n=r-1$.
Then
\begin{eqnarray}
A_{\omega}(\ell )&=&-\frac{t_{0}}{t}\frac{g_{\omega}(\ell
,\ell)a_{\omega}^{-r}-g_{\omega}(r,\ell
)a_{\omega}^{-\ell}}{a_{\omega}^{-R}-a_{\omega}^{R}},\nonumber\\
B_{\omega}(\ell )&=&-\frac{t_{0}}{t}\frac{g_{\omega}(r
,\ell)a_{\omega}^{\ell}-g_{\omega}(\ell ,\ell
)a_{\omega}^{r}}{a_{\omega}^{-R}-a_{\omega}^{R}}.
\end{eqnarray}
Finally we write Eq. (\ref{geom}) for $n=\ell ,r$ and obtain
\begin{eqnarray}
g_{\omega}(r,r)=g_{\omega}(\ell ,\ell )&=&\frac{1}{2}\Biggl
(\frac{1}{D_{\omega}^{+}}+\frac{1}{D_{\omega}^{-}}\Biggr ),\nonumber\\
g_{\omega}(\ell,r)=g_{\omega}(r,\ell )&=&\frac{1}{2}\Biggl
(\frac{1}{D_{\omega}^{+}}-\frac{1}{D_{\omega}^{-}}\Biggr ),\label{gdn}
\end{eqnarray}
where
\begin{eqnarray}
D_{\omega}^{\mp}&=&\omega-\epsilon_{0}+\frac{t_{0}^{2}}{t}a_{\omega}
\nonumber\\
&+&\frac
{t_{0}^{2}}{t}\frac{a_{\omega}^{-R+1}-a_{\omega}^{R-1}}{a_{\omega}^
{-R}-a_{\omega}^{R}}\pm\frac{t_{0}^{2}}{t}\frac{a_{\omega}-a_{\omega}
^{-1}}{a_{\omega}^{-R}-a_{\omega}^{R}}\nonumber\\
&=&\omega -\epsilon_{0}+\gamma \Bigl
(a_{\omega}+\frac{a_{\omega}^{\frac{R}{2}-1}\mp
a_{\omega}^{1-\frac{R}{2}}}
{a_{\omega}^{\frac{R}{2}}\mp
a_{\omega}^{-\frac{R}{2}}}\Bigr ).\label{DPM}
\end{eqnarray}

\begin{figure}
\centerline{\psfig{figure=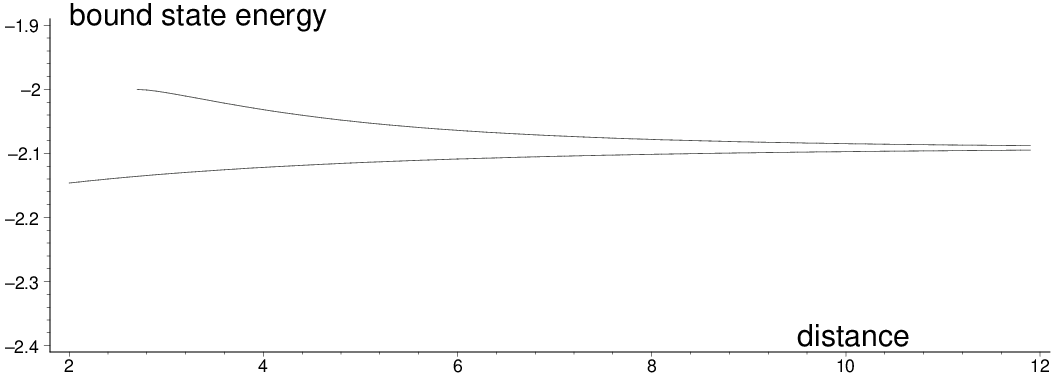,width=2.25cm,height=12cm}}
\vspace{-5cm}
\caption{The two single-particle bound state energies below the band, as
function of the distance between the impurities.}
\label{2imp}
\end{figure}

The single-particle bound energies are determined by the poles of the
Green's functions, i.e., when $D_{\omega}^{\pm}$ vanishes. Let us for
simplicity confine ourselves to bound states below the band, with
$0<a_\omega \le 1$. Then
$D_{\omega}^{+}$ produces a bound state with energy 
$\epsilon_{\beta +}$
as long as
$\epsilon_{0}<2(\gamma -1)$, as is the case for the single impurity
configuration. However, the second bound state, $\epsilon_{\beta -}$,
coming from
$D_{\omega}^{-}$, appears only at more negative $\epsilon_0$,
or (for fixed $\epsilon_0$) when the distance between the
impurities, $R$, is large enough: $\epsilon_{0}<2(\gamma
-1-\gamma /R)$ (solve $D_\omega^-=0$ with $a_\omega \rightarrow 1$
and $\omega=-2$).
As $R$ tends to $\infty$, the two bound energies are
approaching the same value, that of the bound energy of the single
impurity system. We exemplify this behavior in Fig. \ref{2imp}, for
$\gamma =0.4$ and $\epsilon_{0}=-1.5$; the state with the higher energy
appears only for $R>R_c \equiv \gamma/(\gamma-1-\epsilon_0/2)=2.666...$.
The calculations presented in this paper are also restricted to
$\epsilon_0<2(1-\gamma)$, so that there exist no bound states above the
band.

Generally, all the eigenstates of the problem divide into two
subsets. Those which arise from $D_\omega^\pm=0$ obey
$g_{\omega}(\ell,r)/g_{\omega}(r,r)=\pm 1$, and hence represent
even (odd) solutions which obey
\begin{equation}
g_\omega^\pm(\ell-m,\ell)=\pm g_\omega^\pm(r+m,\ell).
\end{equation}
In particular, one can associate these subsets with the even and odd
single-electron wave functions, $\phi_a(\ell-m)=\pm \phi_a(r+m)$.
The two bound states below the band, $\phi_{\beta \pm}(n)$, thus
correspond to the ``bonding" and ``antibonding"
states of molecular chemistry. The antibonding energy $\epsilon_{\beta
-}$
joins the band
for inter-impurity distances below $R_c$.

\begin{references}
\bibitem[*]{OEW} Permanent address: School of Physics and Astronomy,
Raymond and Beverly Sackler Faculty of Exact Sciences, Tel Aviv
University, Tel Aviv 69978, Israel.

\bibitem{CB} G. Grabert and M. H. Devoret, eds., {\it Single Charge
Tunneling, Coulomb Blockade Phenomena in Nanostructures}, NATO ASI,
Series B Physics, vol.
{\bf 294}, Plenum, NY (1992); L. P. Kouwenhoven {\it
et al.}, in {\it Mesoscopic Electron Transport}, edited by L. L. Sohn,
{\it et al.,} (Kluwer, Dordrecht, The Netherlands, Series E, vol. {\bf
345}, 105, 1997).

\bibitem{K} M. A. Kastner, Rev. Mod. Phys. {\bf 64}, 849 (1992).

\bibitem{M} K. V. Matveev, Phys. Rev. B{\bf 51}, 1743 (1995);
A. Furusaki and K. A. Matveev, Phys. Rev. Lett. {\bf 75}, 709 (1995).

\bibitem{zhitenev} N. B. Zhitenev, R. C. Ashoori, L. N. Pfeiffer, and
K. W. West, Phys. Rev. Lett. {\bf 79}, 2308 (1997).

\bibitem{tarucha} S. Tarucha, D. G. Austing, Y. Tokura, W. G. van der
Wiel and
L. P. Kouwenhoven, Phys. Rev. Lett. {\bf 84}, 2485 (2000).

\bibitem{ng}T. K. Ng and P. A. Lee, Phys. Rev. Lett. {\bf 61}, 1768
(1988).

\bibitem{weprb} A. Aharony, O. Entin-Wohlman, and Y. Imry, Phys. Rev.
B{\bf 61}, 5452 (2000).

\bibitem{epl} O. Entin-Wohlman, A. Aharony, Y. Imry and Y. Levinson,
Europhys. Lett. {\bf 50}, 354 (2000).

\bibitem{das} X. Hu and S. Das Sarma, quant-ph/9911080.

\bibitem{Fuji} T. Fujisawa {\it et al.}, Science {\bf 282},
932 (1988).

\bibitem{oppen} F. von Oppen, T. Wettig, and J. M\"{u}ller, Phys. Rev.
Lett. {\bf 76}, 491 (1996); M. Ortu$\tilde{n}$o and E. Cuevas,
Europhys. Lett. {\bf 46}, 224 (1999).

\bibitem{foot}  This nomenclature is based on the picture of  
dielectronic donors in a
lightly doped semiconductor: having only a single-electron 
bound state on the donor would mean
an ionization of one electron into the conduction band
at the transition. 

\bibitem{exchange} P. W. Anderson, Phys. Rev. {\bf 115}, 2 (1959).

\bibitem{tarucha1} Such an interchange was discussed, 
in a model without
leads, in Ref. \onlinecite{tarucha}.

\bibitem{harrison} The same crossover was discussed, for the case without
leads, by W. A. Harrison, Phys. Rev. B{\bf 29}, 2917 (1984).

\end{references}
\end{document}